\documentclass[12pt]{iopart}
\usepackage{graphics}
\usepackage{epsfig}
\begin{document}

\title{Threshold Resonant Structure of the
 $^{232}$Th Neutron-Induced Fission Cross Section}

\author{M. Mirea\dag\  L. Tassan--Got\ddag\ C. Stephan\ddag\ C.O.Bacri\ddag\ 
 and R.C. Bobulescu$^{1}$}

\address{\dag\ Horia Hulubei National Institute for Physics and Nuclear
Engineering, 077125 Bucharest, Romania}

\address{\ddag\ Institute de Physique Nucleaire, 91406 Orsay Cedex, France}

\address{$^{1}$ Faculty of Physics, P.O. Box MG-11, Bucharest, Romania}

\begin{abstract}
The structures observed in the sub-threshold  neutron-induced
fission of $^{232}$Th were investigated employing a recent 
developed model. Theoretical single-particle excitations of a
phenomenological two-humped barrier are
determined by solving a system of coupled differential equations for the motion
along the optimal fission path.  A rather good agreement with experimental
data was obtained using a small number of independent parameters.
It is predicted that the structure at 1.4 and 1.6 MeV is mainly dominated
by spin 3/2 partial cross-section with small admixture of spin 1/2,
while the structure at 1.7 MeV is given by a large partial cross
section of spin 5/2.

\end{abstract}


\maketitle

\section{Introduction}
\label{intro}

The measured neutron-induced fission cross-section behavior
of nuclei in the thorium region represented
a challenge for nuclear physicists concerning  the
shape of the potential energy surface.
The experimental data suggested the existence of a triple
humped barrier.
The neutron-induced cross-sections of $^{230,232}$Th exhibit
multiple fine structures \cite{james,blons,blons1} superimposed on
a gross structure of the threshold cross-section. If the
fine structure is interpreted as a serie of rotational states 
constructed on a $\beta$-vibrational state produced in
some well of the deformation energy, it is
straightforward to postulate the existence of a triple-humped
barrier. The spacing between the members of the band is so
small that it is consistent only with a
prolate deformation that reaches the vicinity of the second-barrier
top. The analysis of Ref. \cite{james} indicates that an intermediate
state nucleus must exist at a deformation considerably larger 
that the normal value. A ternary
minimum obtained theoretically in the potential energy surface of
$^{210}$Po \cite{moller1} made this hypothesis credible. 
Therefore, a shallow minimum was assumed at this 
deformation to create a new $\beta$-vibrational state. 
Angular distribution
analysis \cite{blons2,caruana} confirmed the existence of the
triple well.
Up to
now, the assumption of a triple-humped barrier seems to be
the best interpretation for the fine structure of intermediate
cross-section resonances \cite{boldeman}.  

On the other hand, our analysis explores a different way to consider
the cross-section resonant structure phenomenon by quantifying
the dynamical single-particle effects associated to vibrational
resonances produced in the second well \cite{mir1}.
Our exploratory investigation showed that the $^{230}$Th neutron-induced fission
threshold resonant structure can be explained
\cite{mir2} by rearrangements of single-particle orbitals on the way from the
initial configuration of the compound nucleus up to scission. This
resonant structure depends also on the dynamics of the process.

Sect. \ref{sec2} will provide a general description of the
formalism intended for the evaluation of single-particle excitations, 
while results concerning the intermediate structure of the fission cross-section
will be extensively presented in
Sect. \ref{sec3}. Comments are made in Sect. \ref{sec4}.

\section{Single-particle excitations}
\label{sec2}
In most usual theoretical treatments of nuclear fission, the whole
nuclear system is characterized by some collective coordinates
associated with some degrees of freedom that determine 
approximately the behavior of many other intrinsic variables.
The basic ingredient in such an analysis is a shape 
parametrization that depends on several macroscopic
degrees of freedom. The generalized coordinates associated
to these degrees of freedom vary in time leading to
a split of the nuclear system in two separated fragments.
A microscopic potential must be constructed, to be consistent 
with this nuclear 
shape parametrization. The three important degrees of
freedom encountered in fission, that is, elongation, necking and
mass-asymmetry, must be taken into account.
By solving the Schr\"{o}dinger equation for a reasonable
mean field potential associated to the nuclear shape
parametrization, 
the single-particle energies are determined.
In the case of odd-nucleon systems, the potential barrier
must be increased with an excitation associated to the
unpaired nucleon. The amount of which the barrier
is increased can be estimated within the specialization
energy \cite{wheeler}. This quantity can be interpreted 
as the excess of the energy of the unpaired
nucleon with a given spin over the energy of the same spin nucleon
state of lowest energy.

\begin{figure}

\resizebox{0.45\textwidth}{!}{%
  \includegraphics{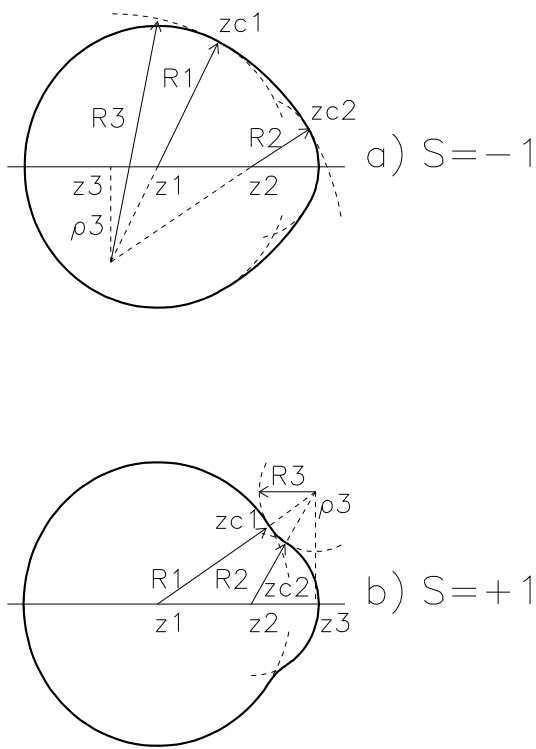}
}
\caption{Nuclear shape parametrization. $z_{1}$, $z_{2}$ and $z_{3}$ 
are the positions
of the centers of circles of radii $R_{1}$, $R_{2}$ characterizing the 
two nascent
fragments,  and of $R_{3}$ determining the neck, respectively. 
If $s=1$, the shape is necked, otherwise the shape is
swollen in the median surface. The distance between the
two centers $z_{1}$ and $z_{2}$ determines the elongation $R$.
}
\label{fig1}       
\end{figure}

In the present work,
an axial-symmetric nuclear parametrization is obtained by smoothly joining two
intersected spheres of different radii $R_{1}$ and $R_{2}$ with
a neck surface generated by the rotation  of a circle of radius
$R_{3}$ around the symmetry axis, as displayed in Fig. \ref{fig1}.
The surface equation is given in cylindrical coordinates:
\begin{equation}
\rho_{s}(z)=\left\{\begin{array}{l}
\sqrt{R_{1}^{2}-(z-z_{1})^{2}},~~~z\le z_{c1}\\
\rho_{3}-s\sqrt{R_{3}^{2}-(z-z_{3})^{2}}, ~~~z_{c1}<z<z_{c2}\\
\sqrt{R_{2}^{2}-(z-z_{2})^{2}},~~~z_{c2}\le z,
\end{array}
\right.
\label{eq1}
\end{equation}
where $z_{c1}$ and $z_{c2}$ define the region of the necking.
The meaning of the geometrical symbols that depends on the shape 
parametrization can be understood inspecting Fig. \ref{fig1}. This parametrization allows to
characterize a single nucleus or two separated nuclei. 
Throughout the paper, the subscripts 0, 1, and 2
indicate the parent, the heavy and light fragments, respectively.
If $S$=1, the shapes are necked in the median surface
characterizing scission shapes and
if $S$=-1 the shapes are swollen characterizing the ground-state and saddle points.
The macroscopic parameters used in the following are
denoted $R=z_{2}-z_{1}$ (elongation), $C=S/R_{3}$ (necking)
and $\eta=R_{1}/R_{2}$ (mass-asymmetry).
For large distances between the two nascent fragments, the
configuration given by two separated spheres is reached.

For instance, to minimize the action integral \cite{leder} 
it is very difficult to treat the three independent generalized coordinates
in the same time. Some simplifying assumptions must be introduced.
As mentioned also in Ref. \cite{bocquet}, microscopic approaches to fission 
\cite{berger,moller}
established that the second saddle point is asymmetrical with a value
compatible with the observed mass ratio. 
In the same time, in the region of the second barrier,
the mass-asymmetry component of the inertia tensor is very large
\cite{mnap}. So, the variations of the mass-asymmetry coordinate
are hindered in this region. On another hand, for elongation smaller
than that of the second well deformation, the mass-asymmetry component of
the inertia is much lower. Therefore, the mass-asymmetry coordinate 
can be modified without enhancing too much
the value of the action integral. Moreover, the deformation energy
is less sensitive to variations of the mass-asymmetry coordinate
in the region of compact shapes.
As in Ref. \cite{mnap}, this observation
allows us to reduce the number of parameters in order to
rend our problem tractable. Therefore, the evolution of the mass
asymmetry generalized coordinate will be a priori fixed in the following. 
It is assumed that the ratio $R_{1}/R_{2}$ varies
linearly from unity (first barrier top) to the value
associated with the final mass partition (second barrier top).
The mass asymmetry in the outer barrier region is deduced
by considering that the volume occupied by the light 
fragment equals the final one. 

\begin{figure}
\resizebox{0.60\textwidth}{!}{%
  \includegraphics{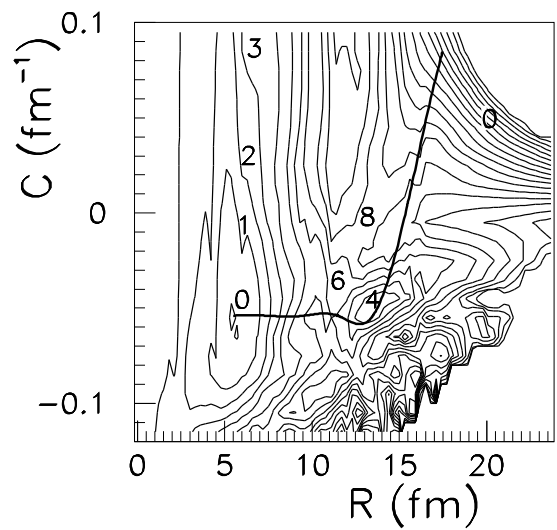}
}
\caption{  Deformation energy in MeV for the
partition $^{233}$Th$\rightarrow^{98}$Sr$+^{135}$Te. $C$ represents
the curvature of the neck and $R$ the distance between the centers of
the fragments. Positive values of $C$ characterize necked-in shapes.
The mass-asymmetry is varied linearly with
$R$ from a value $\eta(R\approx 5~{\rm  fm})$=0 (close to the
ground-state of the compound nucleus) to the final value
$\eta=A_{1}/A_{2}$ (in the vicinity of the top
of the second barrier). The step between two equipotential lines
is 1 MeV. Several values of the deformation energy are marked on the
plot. The dynamic trajectory is represented with a thick
line that starts in the first well, penetrates the first barrier,
attains the second well and finally tunnels the second barrier towards
scission. 
}
\label{fig2}
\end{figure}

The deformation energy of the nuclear system
is the sum between the liquid drop energy
and the shell effects, including pairing corrections. 
The macroscopic energy is obtained in
the framework of the Yukawa-plus-exponential model extended for
binary systems with different charge densities \cite{poenaru}.
The Strutinsky prescriptions \cite{brack} were computed on the basis of
the Superasymmetric Two Center Shell Model (STCSM) \cite{mi1,mi2}.
For one of the most probable partition 
$^{233}$Th $\rightarrow$ $^{98}$Sr $+^{135}$Te, the
deformation energy as function of $C$ and $R$ is plotted in Fig. \ref{fig2}.

The theoretical study of binary disintegration processes
is limited by the difficulties encountered in the calculation
of single-particle levels for very deformed one-center
potentials. On one hand, central oscillator potentials are not
able to describe in a correct manner the shapes for the passage
of one nucleus to two separated nuclei without including a large number
of multipole deformation parameters and, on the other hand,
for very large prolate deformations the sum of single-particle energies
reaches an infinite value, as evidenced within the deformed oscillator
model. These difficulties are surpassed by
considering that the mean field is generated by nucleons
moving in a double center potential. This kind of models allows to describe
scission configurations within a small number of degrees of freedom.
A more realistic version
of the two-center shell model 
was realized recently \cite{mi2} and it is used to generate the
single-particle energy evolutions from the ground-state up to
the formation of two separated fragments. The nuclear shape
parametrization being characterized by an axial symmetry, the
good quantum numbers are the projection of the spin $\Omega$.
As in the Nilsson
model, the single-particle energies depend on two interaction
constants $\kappa$ and $\eta$,
related to the spin-orbit operator and to the squared orbital momentum
correction, respectively. These constants are determined in order to reproduce
the ground-state properties \cite{nilsson}.

In order to determine the single-particle excitations, it is not
sufficient to have a model for the intrinsic nuclear levels,
but is necessary to perform a full calculation of the
trajectory of the decaying system in the configuration space.
The shape of the fission barrier can be obtained if the
trajectory of the nuclear system in our three-dimensional
configuration space is obtained, starting with the ground-state
of the compound nucleus and reaching the exit from the second
barrier. 
This trajectory can be obtained by minimizing numerically
the action functional that gives the quantum penetrability:
\begin{equation}
P=\exp\left\{-{2\over\hbar}\int_{R_{i}}^{R_{f}}
\sqrt{ 2V(R,C,\eta)M\left(R,C,\eta,{\partial C\over\partial R},
{\partial\eta\over\partial R}\right)}dR\right\}
\label{wkb}
\end{equation} 
in the semi-classical Wentzel-Kramers-Brillouin
approximation \cite{brack}. The two turning points
$R_{i}$ and $R_{f}$ denote the elongations that
characterize the ground-state and the exit point of the
barrier, respectively.
Here $V(R,C,\eta)$ is the deformation energy and 
$M(R,C,\eta,{\partial C\over\partial R},{\partial\eta\over\partial R})$ is
the effective mass along the trajectory. The inertia is computed
in the frame of the Werner-Wheeler approximation \cite{davies},
that means, the flow of the fluid is idealized as non-rotational,
non-viscous and hydrodynamic. Using the minimal action principle,
in general, the nuclear system does not follow a path characterized
by minimal values of the deformation energy, so that the trajectory
does not interpolate barrier saddle points values.

\begin{figure}
\resizebox{0.60\textwidth}{!}{%
  \includegraphics{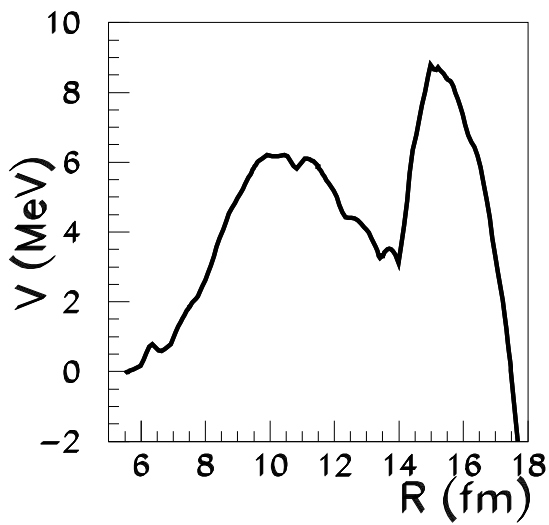}
}
\caption{ Theoretical dynamical barrier calculated along the minimal action
trajectory as function of the elongation $R$.}
\label{fig3}
\end{figure}

\begin{figure}
\resizebox{0.60\textwidth}{!}{%
  \includegraphics{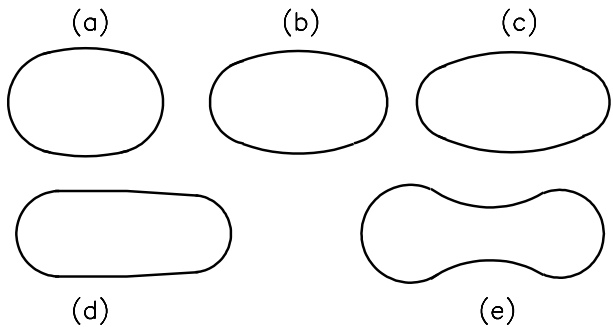}
}
\caption{ The shapes obtained along the minimal action trajectory.
(a) The ground state with elongation $R$=5.8 fm and necking coordinate
$C$=-0.053 fm$^{-1}$. (b) The region of the first barrier 
with $R$=10.57 fm and 
$C$=-0.04 fm$^{-1}$. (c) The region of the second well with
$R$=13.69 fm and 
$C$=-0.0508 fm$^{-1}$. (d) The region of the second barrier with
$R$=15.139 fm and 
$C$=-0.008 fm$^{-1}$. (e) The region of the exit from the barrier
with $R$=17 fm and 
$C$=0.085 fm$^{-1}$.}
\label{shapes}
\end{figure}

Having in mind the assumption imposed for the variation of
the mass-asymmetry, 
the action integral must be minimized in a two-dimensional
space spanned by $C$ and $R$. The first turning point $R_{i}$
is fixed but the second $R_{f}$ lies on the equipotential
line that characterizes the exit from the outer barrier,
that is $R_{f}$ is a function of $C$.
A simple numerical method is used to find the paths
characterized by different values of $R_{f}$, associated
with local minimums. For that purpose, the function
$C=f(R)$ is approximated with a spline function of $n$ variables
$C_{j}$ $(j=1,n)$ in fixed mesh points $R_{j}$ located
in the interval $[R_{i},R_{f}]$ along the elongation axis.
A numerical expression for the WKB functional (\ref{wkb})
that depends only on the parameters $C_{j}$ is obtained.
This expression is minimized numerically. For every
value of $R_{f}$ a local minimum is obtained. The best
values are retained. The trajectory is displayed on
Fig. \ref{fig2}. This dynamical trajectory starts from the
ground-state, reaches the region of the second well and
the slope changes suddenly to penetrate the outer barrier.
Between the first and second well, the macroscopic coordinate
$C$ is less than 0, that is the shapes are swollen in the median region.
Penetrating the second well, the shapes become necked.
The theoretical potential barrier obtained along
the minimal action path is plotted in Fig. \ref{fig3}.
The height of the outer barrier is very large, therefore
some corrections are required in order to obtain realistic
values of the fission cross-section. This is the main reason
that leads to use a phenomenological barrier in calculating
the cross section. The first well is located 
at approximately R=5.5 fm and identifies the fundamental state.
In Fig. \ref{shapes}, the nuclear shapes of the extreme values
of the barrier are displayed.

\begin{figure*}
\resizebox{0.75\textwidth}{!}{%
  \includegraphics{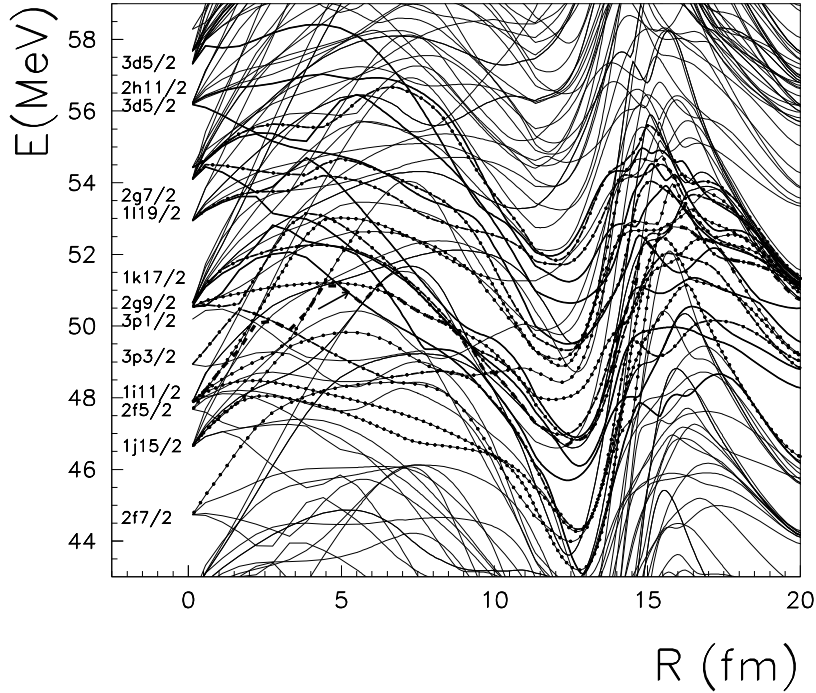}
}
\caption{Neutron level scheme as function of the elongation.
At elongation zero, the shape parametrization gives a spherical
nucleus and the spectroscopic notations are available.
For low values of the deformations, the system behaves as
a Nilsson level scheme. Asymptotically ($R\rightarrow\infty$)
the two diagram of the two formed fragments are superimposed.
In the adiabatic representation, the last occupied level
is displayed with a thick dashed line. The 8 selected levels with
$\Omega$=1/2 are represented with full thick line, the 5 levels
with $\Omega$=3/2 are plotted with dot-dashed thick lines,
the 4 times $\Omega$=5/2 and 3 times 7/2 levels are marked with dotted lines 
(smaller distance between points for $\Omega$=5/2).
The ground-state of the compound nucleus is indicated with an
arrow.   
}
\label{fig4}
\end{figure*}

Using the STCSM the neutron diagram is computed along the minimal action
trajectory, as displayed in Fig. \ref{fig4}. Up to
$R\approx$ 5.5 fm the nuclear system is considered 
reflection symmetric. From the
ground-state (located at approximately $R$=5.5 MeV) up to scission,
the system loses the reflection symmetry to reach the final
partition $^{233}$Th$\rightarrow^{98}$Sr$+^{135}$Te. In these
circumstances, the parity is no longer a good quantum number,
the levels being characterized only by the spin projection
$\Omega$ as good quantum numbers. 
The Nilsson coefficients of the orbital momentum operators ($\kappa$=0.063
and $\eta$=0.8)
were determined to reproduce as better as possible
the experimental sequence of the first excited levels in $^{233}$Th.
The first single-particle excited states are retrieved: 
an ${1\over 2}^{+}$ state (fundamental level) emerging from 
2g$_{9/2}$ followed by a ${5\over 2}^{+}$ one.

\begin{figure}
\resizebox{0.75\textwidth}{!}{%
  \includegraphics{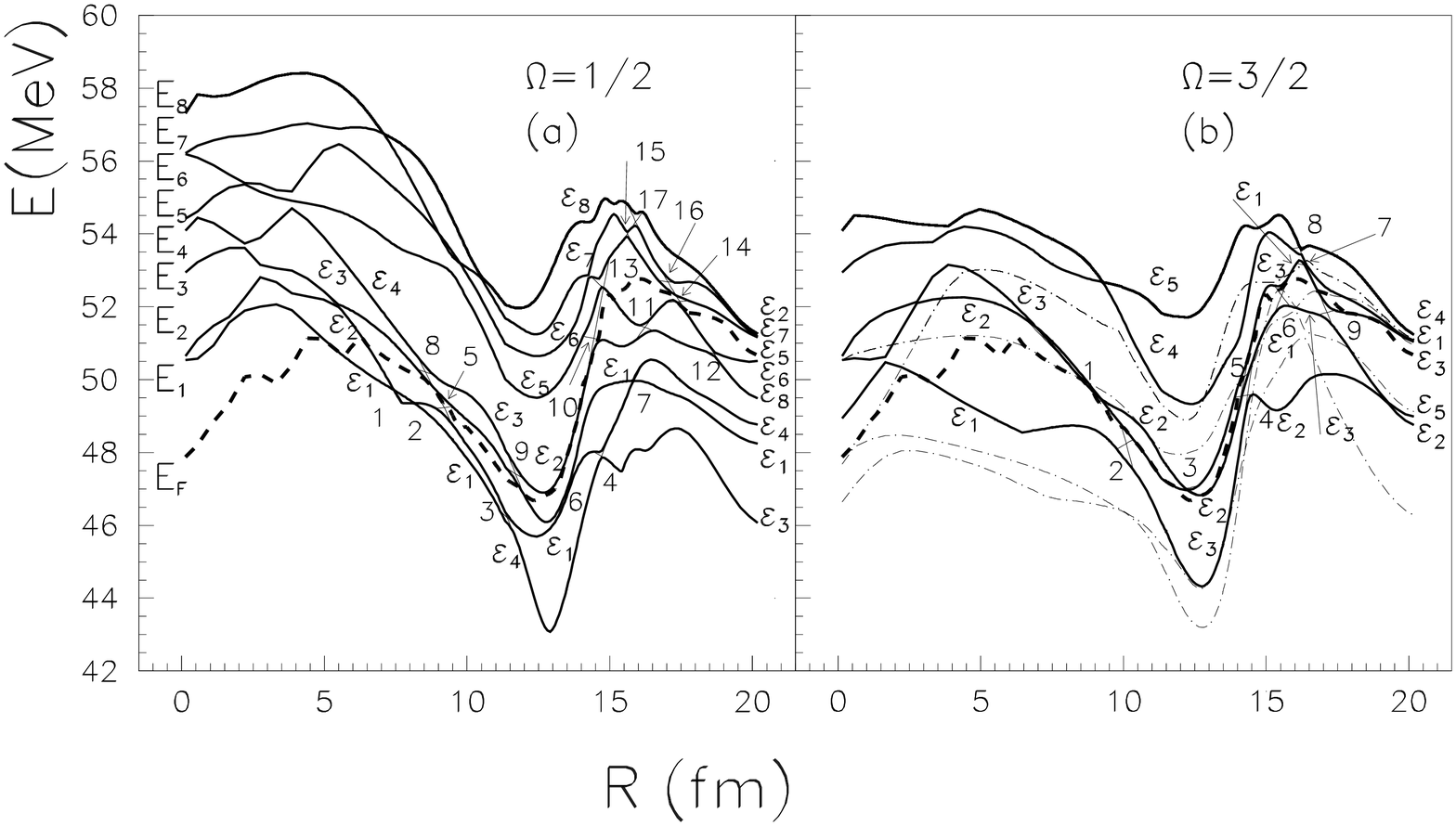}
}
\caption{(a) The 8 selected levels with $\Omega$=1/2.
The last occupied level in the adiabatic representation
is denoted $E_{F}$ and is represented with a dashed line. 
The avoided level crossing regions
are numbered and the diabatic levels $\epsilon_{i}$ identified.
In the ground-state configuration, the $\epsilon_{1}$ level
(emerging from $E_{1}$) is superimposed to $E_{F}$.
At $R\approx$ 20 fm, $E_{F}$ is located between $\epsilon_{6}$
and $\epsilon_{5}$, while the adiabatic level emerging from
$E_{1}$ dropped to $\epsilon_{3}$.
(b) As in plot (a) for the 5 levels with $\Omega$=3/2. With
thin dot dashed lines the 4 $\Omega$=5/2 adiabatic levels are
also displayed.}
\label{fig5}
\end{figure}

To determine the cross section, several single-particle levels
are selected that lie as close as possible to the Fermi energy
region. These levels give the major contribution in the strength of the
fission channel due to their low excitation energy and the large
amount of macroscopic kinetic energy available for fission.
Concerning the $\Omega$=1/2 workspace, 8  selected levels, 
$E_{1}$ up $E_{8}$  are
extracted separately in the left panel of Fig. \ref{fig5} as an example. 
The last occupied level is denoted $E_{F}$. The diabatic levels 
of the subspace $\Omega=3/2$ are displayed in the right panel
of the same figure. In the following, for simplicity,
the discussion will be restricted only for the subspace
$\Omega={1\over 2}$. For $\Omega={3\over 2},{5\over 2},{7\over 2}$, the
same procedure as in the case of $\Omega={1\over 2}$ will be used.

\begin{figure}
\resizebox{0.50\textwidth}{!}{%
  \includegraphics{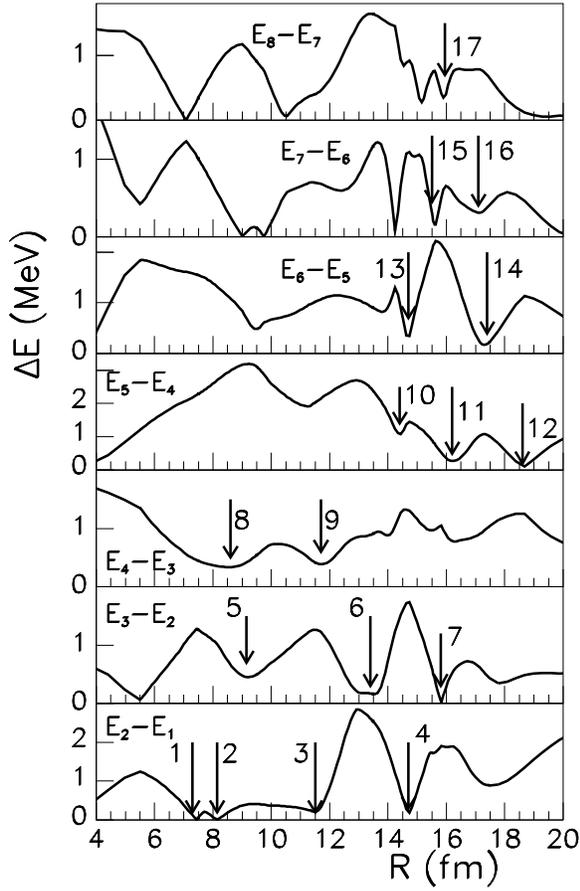}
}
\caption{Differences between the selected adiabatic levels.
The avoided level crossing regions that appears between the
adiabatic energies emerging from the
initial states $E_{1},...E_{4}$ are numbered as in Fig. \ref{fig5}.
}
\label{fig6}
\end{figure}

\begin{figure}
\resizebox{0.50\textwidth}{!}{%
  \includegraphics{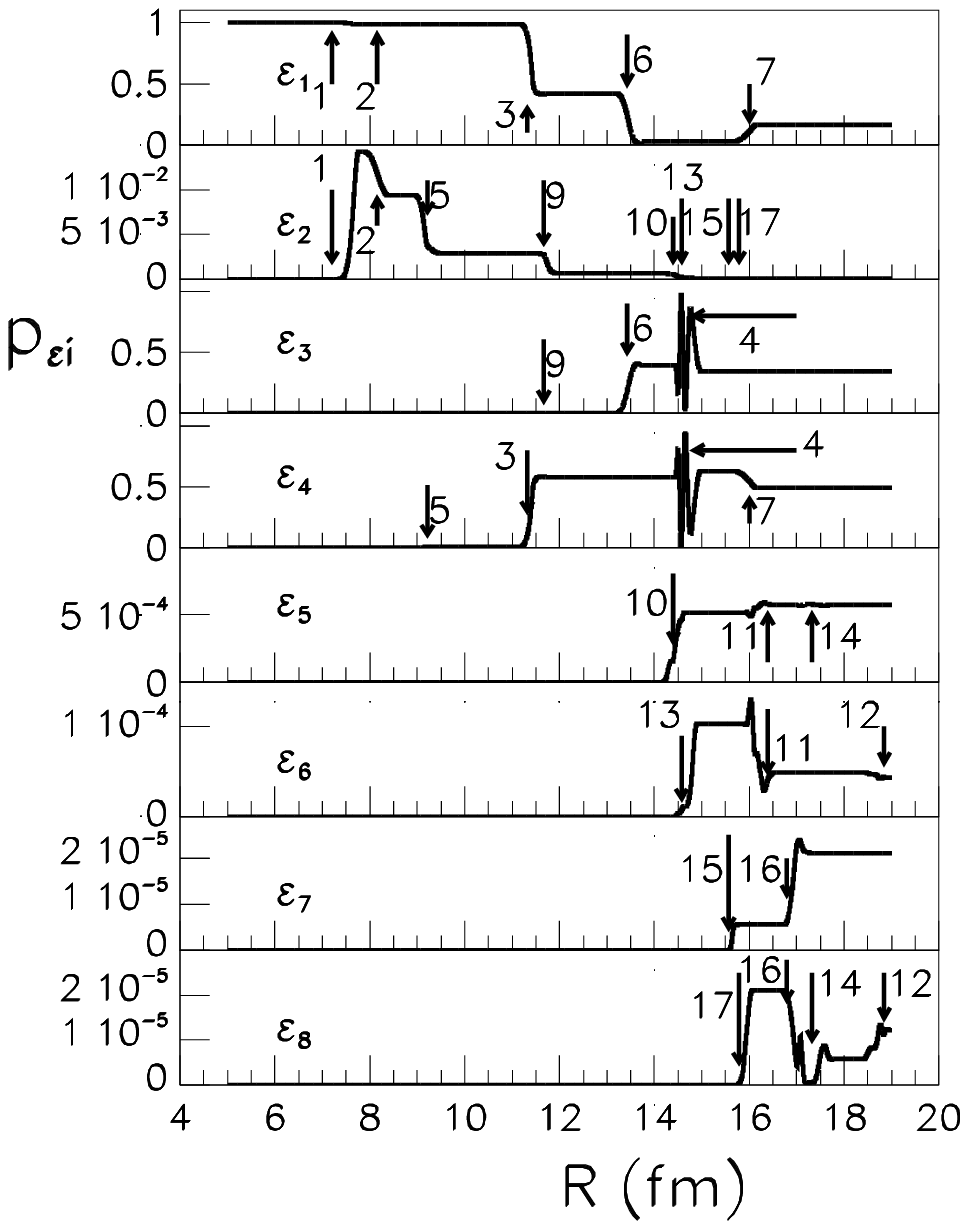}
}
\caption{The occupation probabilities of the diabatic
levels $\epsilon_{1},...\epsilon_{8}$  as function of
the distance between the centers of the fragments.
The same numbers as in Fig. \ref{fig6} are used to
identify the avoided level crossing regions.
The internuclear velocity is 3.5$\times 10^{4}$ m/s,
that leads to a reasonable reaction time 
(time to penetrate the barrier) of approximately
5$\times 10^{-19}$ s. This example is constructed for
an initial condition $p_{\epsilon_{1}}$=1, while $p_{\epsilon_{i}}=0$ 
($i\ne 1$). The occupation probabilities vary in the avoided
level crossing regions.
}
\label{fig7}
\end{figure}

A first behavior can be noticed. The nucleon located on the
adiabatic level emerging
from $E_{1}$ reaches a very unfavorable energy configuration after 
the scission. In the fundamental state, this unpaired nucleon is located
on the fundamental level 
but arrives, in the adiabatic representation, at several MeV
under the last occupied level (the $\epsilon_{3}$ diabatic level).
So, if the nucleon is initially on the ground-state,
it must follow a diabatic energy path to arrive in a most
favorable energy configuration, that is close
to the last occupied level (in one of the diabatic states
$\epsilon_{6}$, $\epsilon_{5}$, $\epsilon_{7}$ or
$\epsilon_{2}$). So, adiabatically, the fission strength
for states with spin 1/2 is not favored. This effect
is a direct consequence of the rearrangement of low
spin orbitals during the disintegration. The number of levels
with $\Omega={1\over2}$ in the two nascent fragments that
are under the energy of the last occupied level is always 
larger that the same number in the compound nucleus. So,
$\Omega$=1/2 orbitals with larger energies of the parent must 
decrease in energy
to fill the levels located under the Fermi energy
of the two fragments. This aspect somewhat hinders 
the possibility to fission through $\Omega={1\over 2}$ channels.
The next step is to study the energy
paths followed by the unpaired nucleon in the diagram.

The realistic two-center diagram presented before provides
an instrument to study the role of individual orbitals
during the disintegration process in a similar way as the study
of nucleus-nucleus collisions \cite{thiel,park} or the alpha- and
cluster-decays \cite{mirea1,mirea2}. Levels with same
quantum numbers associated to some symmetries of the system
cannot cross during the disintegration process and exhibit
avoided level crossing. In our case, due to the axial-symmetry
of the system, the good quantum numbers are the projection of
the spin $\Omega$. The point of nearest approach between two
levels of same $\Omega$ define an avoided level crossing region.
If the internuclear distance varies, the transition probability of a
nucleon between two adiabatic levels is strongly enhanced in
the avoided level crossing region. This promotion mechanism
is generically called the Landau-Zener effect. 

Concerning the 8 single-particle adiabatic levels ($E_{1},...,E_{8}$) 
belonging to the $\Omega$=1/2
workspace, the first step is to find
the avoided level crossing regions. 
The avoided crossing regions can be obtained by plotting 
the energy differences between these adiabatic levels as in 
Fig. \ref{fig6}. Each pertinent avoided crossing is identified and numbered.
The avoided crossings that have
a chance to be located 
along the diabatic single-particle energy paths emerging from
the lower levels $E_{1}$, $E_{2}$ and $E_{3}$ are considered
pertinent. Due to their low initial excitation energy, the transitions through
these levels carry the major part of the fission strength. That
property allows
to restrict our calculations only for an initial condition in which
the occupation probability of one of these levels is one. The next
step is to determine the probability of
realization of each diabatic energy path emerging from these levels. 
Concerning the $\Omega$=3/2 subspace,
the analysis is realized in the same manner, for initial conditions 
restricted to the first 3 low energy levels.  

Assuming an $n$-state approximation, 
the wave function of the unpaired nucleon
can be formally expanded \cite{greiner} in a basis of $n$ diabatic
wave functions $\phi_{i}(r,R)$ as
\begin{equation}
\Psi(r,R,t)=\sum_{i}^{n} c_{i}(t)\phi_{i}(r,R)\exp\left(-{i\over\hbar}
\int_{0}^{t}\epsilon_{ii} dt\right)
\end{equation}
where the matrix elements with the diabatic states $\phi$
are abbreviated as follows 
\begin{equation}
\epsilon_{ij}=<\phi_{i}\mid H\mid\phi_{j}>
\end{equation}
where $H$ is the STCSM Hamiltonian and $c_{i}$ are amplitudes. Inserting 
$\Psi$ in the time-dependent Schr\"{o}dinger equation,
\begin{equation}
\left<\phi_{i}\mid H-i\hbar{\partial\over\partial t}\mid \Psi\right>=0
\end{equation}
the following system of coupled equations is obtained:
\begin{equation}
\dot{c}_{i}={1\over i\hbar}\sum_{j\ne i}^{n}c_{j}\epsilon_{ij}
\exp\left(-{i\over\hbar}\int_{0}^{t}(\epsilon_{jj}-\epsilon_{ii})dt\right)
\label{sys}
\end{equation}
To solve this system, the internuclear velocity $\dot{R}$, the diabatic 
energies and the interaction matrix elements must be known.
Apart the relative velocity, the other ingredients are 
supplied by the STCSM. 
The diabatic states are constructed by using spline interpolations
in the level crossing regions. The interaction matrix elements
$\epsilon_{ij}$ between the diabatic states is a measure of the
difference between adiabatic and diabatic energies. The occupation
probability of each adiabatic level as function of $R$ is 
now obtained $p_{\epsilon_i}=\mid c_{i}\mid^{2}$. For the unpaired
neutron initially located in the fundamental state $E_{1}$, the
system (\ref{sys}) is solved within the boundary condition 
$c_{1}$=1 and $c_{i}$=0 for $i\ne 1$. The occupation probabilities
of each diabatic level plotted in Fig. \ref{fig5} are represented
in Fig. \ref{fig7}. Within the selected levels and avoided 
level crossings, 40 different energy paths of the unpaired neutron
can be obtained as indicated in Table \ref{tab:1}. Here, an approximation is
made by considering that the points of the avoided level crossings 1 and 2
form a single avoided level region. Otherwise, the number of
paths gets doubled. Each path represents
an excitation of the nuclear system. The probability of each
path can be estimated. For example, it can be deduced from Fig. \ref{fig7}
that the path $\epsilon_{1}-2-\epsilon_{1}-3-\epsilon_{4}$ carries
about 0.5 of the probability. The line between letters and digits
connects diabatic levels and avoided level crossing regions.
A strong mixing is produced in the
region 4, that leads us to conclude that the paths 
$\epsilon_{1}-2-\epsilon_{1}-3-\epsilon_{4}-4-\epsilon_{3}$ (no. 1 in table
\ref{tab:1})
and $\epsilon_{1}-2-\epsilon_{1}-3-\epsilon_{4}-4-\epsilon_{4}-7$
carry each of them about 0.25 probability. Finally,
it can be considered that 
the path 
$\epsilon_{1}-2-\epsilon_{1}-3-\epsilon_{4}-4-\epsilon_{4}-7-\epsilon_{1}$ 
(no. 2) has about 0.05 probability of realization while
$\epsilon_{1}-2-\epsilon_{1}-3-\epsilon_{4}-4-\epsilon_{4}-7-\epsilon_{4}$ 
(no. 3) remains with 0.2. The other probabilities are estimated in the same
manner. The same procedure is repeated for the case
when the unpaired nucleon is initially located on the other
selected levels.

The excitations of the barriers due to one diabatic path $k$
is given by the  specialization
energy. Considering that the fundamental barrier 
corresponds to the nucleon at the Fermi energy, the excitation $E_{x}$
as function of $R$ is 
\begin{equation}
E_{xk}(R)=\sqrt{(\epsilon_{k}(R)-\lambda(R))^{2}+\Delta^{2}(R)}-\Delta(R)
\end{equation}
in the frame of the superfluid model. Here, $\epsilon_{k}$ is the
single-particle energy of the path $k$, $\lambda$ is the Fermi
energy and $\Delta$ the gap. These excitations are added to the
fundamental barrier. These quantities have the same meaning as 
the so-called transition bandheads found in the literature.

\begin{figure}
\resizebox{0.50\textwidth}{!}{%
  \includegraphics{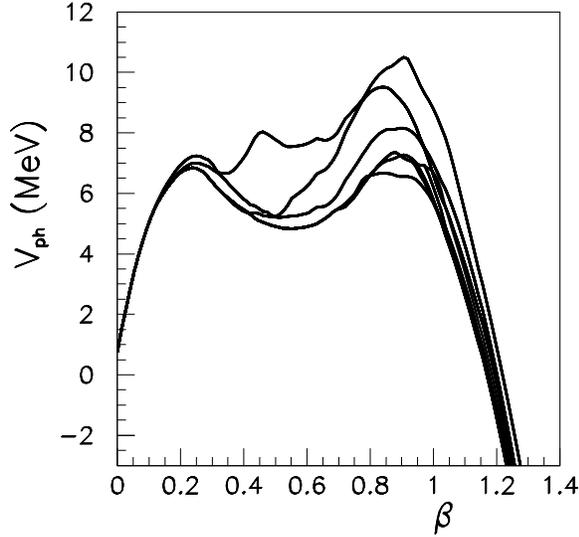}
}
\caption{The $\Omega$=1/2 phenomenological $V_{ph}$
barriers with excitations obtained
in the frame of the hybrid model emerging from the fundamental
level $E_{1}$. 
}
\label{fig8}
\end{figure}

\section{Cross-section}
\label{sec3}

The partial fission cross section $\sigma_{f}$ for a spin $J$ of the compound
nucleus and excitation energy $E^{*}$ is obtained within 
a statistical     principle:
\begin{equation}
\sigma_{f}(J,E^{*})=\sigma_{c}(J,E^{*}){\Gamma_{f}(J,E^{*})\over
\Gamma_{n}(J,E^{*})+\Gamma_{\gamma}(J,E^{*})+\Gamma_{fT}(J,E^{*})}
\end{equation}
where the ratio in the right-hand side is the probability that
the system decays through fission. It is given by a ratio between
energy widths for fission (subscript $f$), neutron emission
(subscript $n$) and $\gamma$-de-excitation (subscript $\gamma$).
The subscript $T$ addresses the total transmission in the fission
channel including absorption in the second well. The neutron transmission
was computed for a squared complex potential \cite{more} in order
to evaluate the compound nucleus cross section. To determine
the participation of different $\Omega$ excitations in the
fission channel for a given spin $J$ of the compound nucleus,
a unfolding procedure in term of Clebsh-Gordon coefficients is used 
\begin{equation}
\begin{array}{c}
\Gamma_{f}(J,E^{*})={1\over 2\pi\rho(J,E^{*},A)}
\sum_{L=0}^{L_{M}}\sum_{\Omega}{<JL\Omega0\mid J\Omega>^{2}\over C}\\
\times \int_{0}^{E^{*}-E_{L}} 2T_{f}(E,L,\Omega)\rho(\Omega,E^{*}-E-E_{L})dE
\end{array}
\end{equation}
where a normalization coefficient is used:
\begin{equation}
C=\sum_{L=0}^{L_{M}}\sum_{\Omega}<JL\Omega0\mid J\Omega>^{2}
\end{equation}
and the condition $J=L+\Omega$ is imposed.
Here  $\rho$ represents the density of states, $E_{L}$ is the
the rotation energy in the fundamental state of the compound nucleus
with an angular momentum $L$ and $L_{M}$ is the maximum orbital momentum
taken into consideration. This formula can be obtained easily
by simplifying the
model underlined in Ref. \cite{mir1}, that is, neglecting the 
additional collective excitations as gamma, sloshing or bending vibrations.
Analog formulas can be obtained for the $\gamma$ and neutron energy widths
as detailed in Ref \cite{mir1}.

In the fission channel, the spin $\Omega$ density of states can be shared
as function of the excitation energy between a discrete component and
a continuum one:
\begin{equation}
\rho(J,E)=\left\{\begin{array}{cc}
\sum_{i}\delta(E-\epsilon_{\Omega,i}),& E<E_{0}\\
\rho_{GC}(\Omega,E), & E\ge E_{0}\end{array}\right.
\end{equation}
where $\rho_{CG}$ is the statistical Gilbert and Cameron 
approximation and $\epsilon_{\Omega,i}$ $(i=1,n)$ are the set of
diabatic single particle energies that are taken into
consideration for a spin projection $\Omega$.
So that, the transmission in the fission channel
can be decomposed as follows:
\begin{equation}
\begin{array}{c}
\int_{0}^{E^{*}-E_{L}} T_{f}(E,L,\Omega)\rho(\Omega,E^{*}-E-E_{L})dE\\
=
\sum_{i}T_{f}(E^{*}-E_{L}-\epsilon_{\Omega,i})\\
+\int_{0}^{E^{*}-E_{L}-E_{0}} T_{f}(E,L,\Omega)\rho(\Omega,E^{*}-E-E_{L})dE
\end{array}
\end{equation}
The sum over $i$  takes into account all the transmissions for
diabatic levels with spin 
$J=\Omega+L$ located in the energy interval $[0,E_{0}]$.
In this context, the transmission $T_{f}(E^{*}-E_{L}-\epsilon_{\Omega,i})$ means a
weighted sum of the transmissions of all available diabatic
energy paths emerging from the level $\epsilon_{\Omega,i}$.

The microscopic model used to compute the theoretical barrier
is subject to some limitations as described in Ref. \cite{mi2}.
It is not possible to obtain pertinent values of the heights
of the barriers. In this circumstances, it is necessary to
use a phenomenological barrier. A phenomenological barrier
is conventionally simulated as a function of a 
dimensionless parameter $\beta$,
that characterizes a deformation, within three smoothed joined
parabolas \cite{cramer}. In our work, an imaginary component of the potential
is added between the turning points of the second well, in order
to simulate the damping due to gamma and neutron emission.
The additional excitations are considered
as specialization energies and are added
to the phenomenological barrier. This operation is  achieved
in the simplest possible way,
by realizing a linear interpolation based on a correspondence
between the elongation $R$ and the dimensionless
parameter $\beta$ in some points. The correspondence was chosen
for the two minimums, the two heights and the exit point.
The hybrid model emerges. New barriers are constructed as
displayed in Fig. \ref{fig8}. 
When only the collective rotations are taken into account, the
heights of the barriers and that of the second well
are modified with a quantity
\begin{equation}
\Delta E_{L}={L(L+2\Omega+1)\hbar^{2}\over 2I_{j}}-E_{L}
\end{equation}
where $I_{j}$ is the moment of inertia, $j$ labels one of
the two heights or the second well. The decoupling parameter is neglected.
The moment of inertia is computed
simply with the formula $I_{j}=\mu R_{j}^{2}$ where $\mu$ is the
reduced mass and $R_{j}$ is the theoretical elongation obtained at the 
extreme point $j$. The quantity 
\begin{equation}
E_{L}={L(L+2\Omega+1)\hbar^{2}\over 2I_{0}}
\end{equation}
addresses the fundamental state of the compound nucleus. The previous formulas
represents an improvement of the formalism found in Ref. \cite{mir1}.

\begin{figure}
\resizebox{0.50\textwidth}{!}{%
  \includegraphics{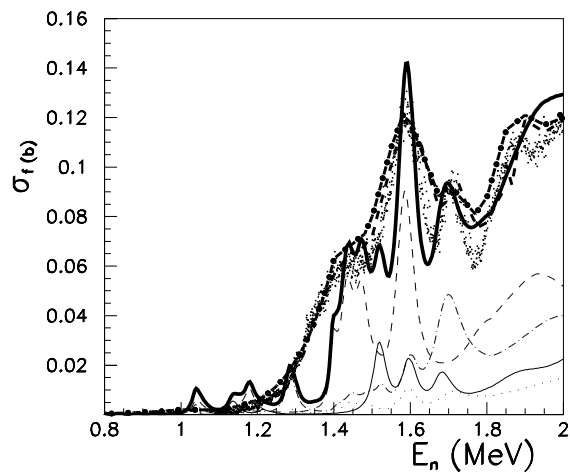}
}
\caption{Thick full-line, neutron-induced cross section for $^{232}$Th
as function of the neutron incident energy $E_{n}$ calculated within
the hybrid model. Points are experimental data. Thick dashed
line represents the ENDF/B-IV evaluation while the thick dot-dashed line
is the JENDL-3.3 one \cite{gov}. Experimental data are from
Ref. \cite{blons,blons1,shc}. A thin full line gives the partial
cross section of spin 1/2, a dashed line is for the spin 3/2,
the dot-dashed one for 5/2 and the dotted line for 7/2.
}
\label{fig9}
\end{figure}

A large number of excited states are obtained that
are characterized by the projection $\Omega$ and the
angular momentum $L$. The transmission
is calculated numerically by approximating
the shape of the excited barrier within 500 constant potential steps using
the numerical recipe found in Ref. \cite{lynn}.
A search of the heights and of the widths
of the phenomenological barrier is realized in order to reproduce
as well as possible the experimental fission cross-section threshold structure.
A behavior that agree
satisfactory with the experimental data is obtained.
The heights of the inner phenomenological barrier,
the second well and the outer barrier are 6.81, 4.83 and 6.61 MeV,
respectively. In the same order, the widths are 1.2, 0.4 and 1.1 MeV.
The theoretical cross-section is represented in Fig. \ref{fig9}
and compared with experimental data and evaluations.

The evaluations succeeded to reproduce better the experimental
data. In general many parameters are taken into account
to evaluate a cross-section in terms of Bohr-channels. For example,
in evaluations
phenomenological level densities functions appropriately
matched to the available experimental structure data at
low excitation energies are used. Multiplication factors are also
applied to level density functions to account for
enhancements in the fission transition state densities
at each fission barrier. It is a common practice to
describe the cross-section as the sum of excitations
for discrete levels constructed to fit the resonance.
In other words, the evaluation takes into account many
other parameters to fit the experimental data 
apart the heights and the widths of the
phenomenological barrier.
In the work presented in this paper, 
no adjustments are made to improve the agreement,
the simulations being based only on the phenomenological
barrier parameters and the internuclear velocity.

Our simulations evidence an oscillatory behavior of the cross-section 
close to 1.4
MeV. This aspect is in agreement with
the experimental data given in Ref. \cite{au}. The experimental data
combined with theoretical arguments estimate a ratio
2:1 between the partial cross section of spin 3/2 and 1/2, respectively.
The model shows that the partial cross section for the spin
3/2 is responsible for the oscillations of the cross section at
these energies. Experimentally, the peak at 1.6 MeV is explained
entirely by a partial cross section of spin 3/2 with a small
5/2 component. In our plot a strong 3/2 component is present
with small admixture of 1/2 and 5/2 partial cross sections.
A discrepancy is obtained for the 1.7 MeV structure. The
experiment evidences the existence of a mixing between
3/2 and 5/2 components while our model predicts a large 5/2 
partial cross section followed by the 3/2 and 1/2 components.

\begin{figure}
\resizebox{0.50\textwidth}{!}{%
  \includegraphics{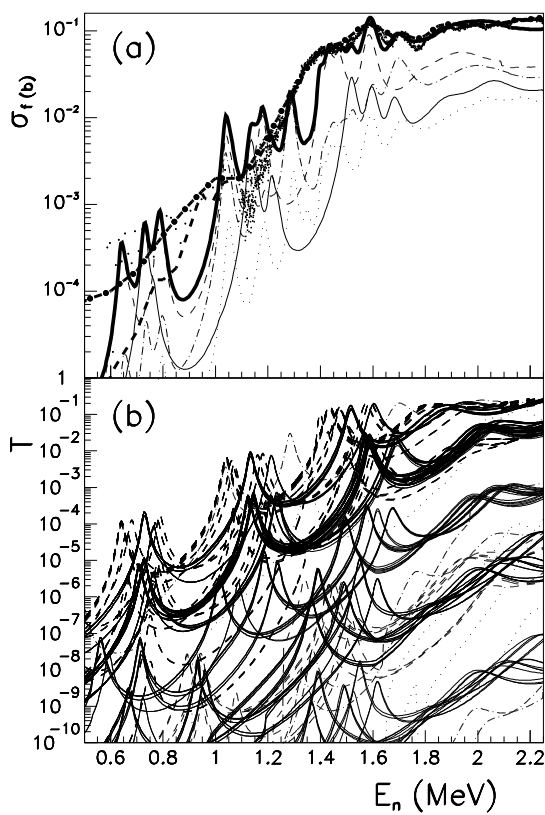}
}
\caption{ (a) Same as Fig. \ref{fig9} in an extended logarithmic scale
along the $y$-axis.
(b) $L$=0 fission transmissions for different barriers
as function of the neutron energy $E_{n}$. 
The transmissions for $\Omega$=1/2 excitations are
plotted with full lines, those for 3/2 with a dashed line,
those for 5/2 with dot-dashed and 5/2 with a dotted line.
}
\label{fig10}
\end{figure}

In Fig. \ref{fig10} the cross-section is plotted on an extended
scale. It can be observed that the theoretical results 
exhibits an oscillatory behavior in the low energy region, 
up to 1.2 MeV, 
around the smooth variation of the experimental data.
In the panel (b), the transmissions computed for the barriers
with different calculated excitations are displayed. The
oscillatory behavior is due to a large number of resonances
associated to the different excited barriers.

\section{Summary and discussion }
\label{sec4}

The scope of the present work is to understand the 
mechanism for the formation of the fission cross
section structure and of the high number of resonances
by appealing essentially to dynamical single-particle effects
associated to $\beta$-vibration in the second well.
The number of free parameters is kept as minimal as possible
(six parameters
that characterize the phenomenological barrier and one
parameter for the internuclear velocity)
to show evidence of the physics of the problem.

Theoretical excitations and their associated probabilities
were determined for a given partition in the
isotopic distribution of fragments. These excitations are added
to a phenomenological barrier in the framework of the hybrid model.
After a suitable search of the parameters of the double humped
phenomenological barrier, the cross section is computed.
The results give a rather good qualitative agreement with experimental
data. 
It is evidenced that the structure at 1.4 and 1.6 MeV is mainly dominated
by spin 3/2 partial cross-section with small admixture of spin 1/2,
while the structure at 1.7 MeV is given by a large partial cross
section of spin 5/2.

In this exploratory analysis, only one partition for the fission
fragments is taken into consideration. For other partitions in the same
mass region,
it is expected that the level scheme changes slightly leading to a small
shift in the energy of the resonances. By taking into account several
partitions in the same mass-region and folding their yields it
it possible to obtain broader resonances as experimentally observed.

In this context, it will be interesting to explore
experimentally if the isotopic fragment distribution in the fission
process changes in the energetic region covered by a resonance,
showing a preference for several mass partitions. If such a phenomenon
can be experimental evidenced, that will represent a strong experimental 
support for our model because the statistical theories don't include
ingredients related to this aspect.

The model can be further improved.
Up to now, only the radial coupling was used to explain the intermediate
structure of the cross-section. It is possible to have better 
results by taking into account the Coriolis mixing and the residual
interactions by using evolved forms for the system of coupled equations
that describes the microscopic motion \cite{mirea2,mirea3}.

Other models succeed to reproduce better the experimental
data \cite{sin} using an extensive number of free parameters:
10 variables for the heights and widths of the triple humped
phenomenological barrier plus 5 times 16 variables for the 
transition bandheads constructed on different intrinsic
excitations (with a significance of 
excitations given by single-particle energies).
Despite the overall excellent agreement on a very large neutron energy
region, this treatment, generally used in evaluations, 
takes into account a peculiar behavior
for the single-particle excitation energies. The levels that
characterize the transition bandheads never intersect.
The first ${1\over 2}^{+}$ level have practically the
same value (having as reference the fundamental state)
during the penetration of the barrier. This behavior,
as remarked previously, cannot be expected. Moreover, the
statistical models consider that the population of each
fundamental transition band is essentially one. The formalism
presented in the Sect. \ref{sec2} indicates that such a behavior 
is physically not reasonable.

The present investigation shows that the resonant structure
of the fission cross section can be explained by the existence
of many barriers associated to single-particle excitations.
So, it is possible that the complex structure in the fission
cross section is due to rearrangement
of orbitals and the dynamic of the process, beginning from
the initial state of the compound nucleus and terminating at the scission.
A large number of different excited barriers are formed
leading to a large number of vibrational resonances in the
second well. These resonances carry information about the structure 
of the nucleus at hyperdeformations and the dynamics. The model
presented in this work represents an alternative to the actual
statistical models and may determine a competitive way 
to consider the fission process.

\begin{table}
\caption{Energy paths open for the first $E_{1}$ $\Omega$=1/2 level 
}
\label{tab:1}  

\footnotesize\rm
\begin{center}    
\begin{tabular}{ l l }
\hline
No. & Energy path \\
\hline
1 & $ \epsilon_{1}-2-\epsilon_{1}-3-\epsilon_{4}-4-\epsilon_{3}$  \\
2 & $\epsilon_{1}-2-\epsilon_{1}-3-\epsilon_{4}-4-\epsilon_{4}-7-\epsilon_{1}$  \\
3 & $\epsilon_{1}-2-\epsilon_{1}-3-\epsilon_{4}-4-\epsilon_{4}-7-\epsilon_{4}$  \\
4 & $\epsilon_{1}-2-\epsilon_{1}-3-\epsilon_{1}-6-\epsilon_{3}-4-\epsilon_{3}$  \\
5 & $\epsilon_{1}-2-\epsilon_{1}-3-\epsilon_{1}-6-\epsilon_{3}-4-\epsilon_{4}-7-\epsilon_{1}$  \\
6 & $\epsilon_{1}-2-\epsilon_{1}-3-\epsilon_{1}-6-\epsilon_{3}-4-\epsilon_{4}-7-\epsilon_{4}$  \\
7 & $\epsilon_{1}-2-\epsilon_{1}-3-\epsilon_{1}-6-\epsilon_{1}-7-\epsilon_{1}$  \\
8 & $\epsilon_{1}-2-\epsilon_{1}-3-\epsilon_{1}-6-\epsilon_{1}-7-\epsilon_{4}$  \\
9 & $\epsilon_{1}-2-\epsilon_{2}-5-\epsilon_{2}-9-\epsilon_{3}-6-\epsilon_{3}-4-\epsilon_{3}$  \\
10 & $\epsilon_{1}-2-\epsilon_{2}-5-\epsilon_{2}-9-\epsilon_{3}-6-\epsilon_{3}-4-\epsilon_{4}-8-\epsilon_{1}$  \\
11 & $\epsilon_{1}-2-\epsilon_{2}-5-\epsilon_{2}-9-\epsilon_{3}-6-\epsilon_{3}-4-\epsilon_{4}-8-\epsilon_{4}$  \\
12 & $\epsilon_{1}-2-\epsilon_{2}-5-\epsilon_{2}-9-\epsilon_{3}-6-\epsilon_{1}-8-\epsilon_{1}$  \\
13 & $\epsilon_{1}-2-\epsilon_{2}-5-\epsilon_{2}-9-\epsilon_{3}-6-\epsilon_{1}-8-\epsilon_{4}$  \\
14 & $\epsilon_{1}-2-\epsilon_{2}-5-\epsilon_{4}-3-\epsilon_{4}-4-\epsilon_{3}$  \\
15 & $\epsilon_{1}-2-\epsilon_{2}-5-\epsilon_{4}-3-\epsilon_{4}-4-\epsilon_{4}-8-\epsilon_{1}$  \\
16 & $\epsilon_{1}-2-\epsilon_{2}-5-\epsilon_{4}-3-\epsilon_{4}-4-\epsilon_{4}-8-\epsilon_{4}$  \\
17 & $\epsilon_{1}-2-\epsilon_{2}-5-\epsilon_{4}-3-\epsilon_{1}-6-\epsilon_{3}-4-\epsilon_{3}$  \\
18 & $\epsilon_{1}-2-\epsilon_{2}-5-\epsilon_{4}-3-\epsilon_{1}-6-\epsilon_{3}-4-\epsilon_{4}-8-\epsilon_{1}$  \\
19 & $\epsilon_{1}-2-\epsilon_{2}-5-\epsilon_{4}-3-\epsilon_{1}-6-\epsilon_{3}-4-\epsilon_{4}-8-\epsilon_{4}$  \\
20 & $\epsilon_{1}-2-\epsilon_{2}-5-\epsilon_{4}-3-\epsilon_{1}-6-\epsilon_{1}-7-\epsilon_{1}$  \\
21 & $\epsilon_{1}-2-\epsilon_{2}-5-\epsilon_{4}-3-\epsilon_{1}-6-\epsilon_{1}-7-\epsilon_{4}$  \\
22 & $\epsilon_{1}-2-\epsilon_{2}-5-\epsilon_{2}-9-\epsilon_{2}-10-\epsilon_{5}-11-\epsilon_{6}-12-\epsilon_{8}$  \\
23 & $\epsilon_{1}-2-\epsilon_{2}-5-\epsilon_{2}-9-\epsilon_{2}-10-\epsilon_{5}-11-\epsilon_{6}-12-\epsilon_{6}$  \\
24 & $\epsilon_{1}-2-\epsilon_{2}-5-\epsilon_{2}-9-\epsilon_{2}-10-\epsilon_{5}-11-\epsilon_{5}-14-\epsilon_{8}-12-\epsilon_{8}$  \\
25 & $\epsilon_{1}-2-\epsilon_{2}-5-\epsilon_{2}-9-\epsilon_{2}-10-\epsilon_{5}-11-\epsilon_{5}-14-\epsilon_{8}-12-\epsilon_{6}$  \\
26 & $\epsilon_{1}-2-\epsilon_{2}-5-\epsilon_{2}-9-\epsilon_{2}-10-\epsilon_{5}-11-\epsilon_{5}-14-\epsilon_{5}$  \\
27 & $\epsilon_{1}-2-\epsilon_{2}-5-\epsilon_{2}-9-\epsilon_{2}-10-\epsilon_{2}-13-\epsilon_{6}-11-\epsilon_{6}-12-\epsilon_{8}$  \\
28 & $\epsilon_{1}-2-\epsilon_{2}-5-\epsilon_{2}-9-\epsilon_{2}-10-\epsilon_{2}-13-\epsilon_{6}-11-\epsilon_{6}-12-\epsilon_{6}$  \\
29 & $\epsilon_{1}-2-\epsilon_{2}-5-\epsilon_{2}-9-\epsilon_{2}-10-\epsilon_{2}-13-\epsilon_{6}-11-\epsilon_{5}-14-\epsilon_{8}-12-\epsilon_{8}$  \\
30 & $\epsilon_{1}-2-\epsilon_{2}-5-\epsilon_{2}-9-\epsilon_{2}-10-\epsilon_{2}-13-\epsilon_{6}-11-\epsilon_{5}-14-\epsilon_{8}-12-\epsilon_{6}$  \\
31 & $\epsilon_{1}-2-\epsilon_{2}-5-\epsilon_{2}-9-\epsilon_{2}-10-\epsilon_{2}-13-\epsilon_{6}-11-\epsilon_{5}-14-\epsilon_{5}$  \\
32 & $\epsilon_{1}-2-\epsilon_{2}-5-\epsilon_{2}-9-\epsilon_{2}-10-\epsilon_{2}-13-\epsilon_{2}-15-\epsilon_{7}-16-\epsilon_{8}-14-\epsilon_{8}-12-\epsilon_{8}$  \\
33 & $\epsilon_{1}-2-\epsilon_{2}-5-\epsilon_{2}-9-\epsilon_{2}-10-\epsilon_{2}-13-\epsilon_{2}-15-\epsilon_{7}-16-\epsilon_{8}-14-\epsilon_{8}-12-\epsilon_{6}$  \\
34 & $\epsilon_{1}-2-\epsilon_{2}-5-\epsilon_{2}-9-\epsilon_{2}-10-\epsilon_{2}-13-\epsilon_{2}-15-\epsilon_{7}-16-\epsilon_{8}-14-\epsilon_{5}$  \\
35 & $\epsilon_{1}-2-\epsilon_{2}-5-\epsilon_{2}-9-\epsilon_{2}-10-\epsilon_{2}-13-\epsilon_{2}-15-\epsilon_{7}-16-\epsilon_{7}$  \\
36 & $\epsilon_{1}-2-\epsilon_{2}-5-\epsilon_{2}-9-\epsilon_{2}-10-\epsilon_{2}-13-\epsilon_{2}-15-\epsilon_{2}-17-\epsilon_{8}-16-\epsilon_{8}-14-\epsilon_{8}-12-\epsilon_{8}$  \\
37 & $\epsilon_{1}-2-\epsilon_{2}-5-\epsilon_{2}-9-\epsilon_{2}-10-\epsilon_{2}-13-\epsilon_{2}-15-\epsilon_{2}-17-\epsilon_{8}-16-\epsilon_{8}-14-\epsilon_{8}-12-\epsilon_{6}$  \\
38 & $\epsilon_{1}-2-\epsilon_{2}-5-\epsilon_{2}-9-\epsilon_{2}-10-\epsilon_{2}-13-\epsilon_{2}-15-\epsilon_{2}-17-\epsilon_{8}-16-\epsilon_{8}-14-\epsilon_{5}$  \\
39 & $\epsilon_{1}-2-\epsilon_{2}-5-\epsilon_{2}-9-\epsilon_{2}-10-\epsilon_{2}-13-\epsilon_{2}-15-\epsilon_{2}-17-\epsilon_{8}-16-\epsilon_{7}$  \\
40 & $\epsilon_{1}-2-\epsilon_{2}-5-\epsilon_{2}-9-\epsilon_{2}-10-\epsilon_{2}-13-\epsilon_{2}-15-\epsilon_{2}-17-\epsilon_{2}$  \\

\end{tabular}
\end{center}
\end{table}

\end{document}